\documentclass[twocolumn,showpacs,amsmath,amssymb]{revtex4}

\usepackage{graphicx}
\usepackage{dcolumn}
\usepackage{bm}
\newcommand{\etal}{\emph{et al.}}
\newcommand{\be}{\begin{equation}}
\newcommand{\ee}{\end{equation}}
\newcommand{\bfig}{\begin{figure}}
\newcommand{\efig}{\end{figure}}
\newcommand{\incl}{\includegraphics}

\begin{document}
\title{High-Field Shubnikov-de Haas Oscillations in the Topological Insulator Bi$_2$Te$_2$Se.
}
\author{Jun Xiong$^1$, Yongkang Luo$^{1,*}$, YueHaw Khoo$^1$, 
Shuang Jia$^2$, R. J. Cava$^2$ and N. P. Ong$^1$
}
\affiliation{
Departments of Physics$^1$ and Chemistry$^2$, Princeton University, Princeton, NJ 08544
}

\date{\today}
\begin{abstract}
We report measurements of the surface Shubnikov de Haas oscillations (SdH) on 
crystals of the topological insulator Bi$_2$Te$_2$Se. In crystals with large bulk resistivity
($\sim$4 $\Omega$cm at 4 K), we observe $\sim$15 surface SdH 
oscillations (to the $n$ = 1 Landau Level) in magnetic fields $B$ up to 45 Tesla. 
Extrapolating to the limit $1/B\to 0$, we confirm the $\frac12$-shift expected from a Dirac spectrum.
The results are consistent with a very small surface Lande $g$-factor.
\end{abstract}


\maketitle                   
\section{Introduction}
In Topological Insulators, the surface electrons occupy helical
Dirac states in which the spin is locked perpendicular to the momentum~\cite{Fu07,FuKane07,Moore07,Bernevig06}. 
In three-dimensional examples, the topological surface state was observed by angle-resolved photoemission spectroscopy (ARPES)~\cite{Hsieh08,Hsieh09,Xia09,Chen09}. 
Scanning tunneling microscopy (STM) has also been applied extensively~\cite{Pedram09,Hanaguri,Xue10}. 
In transport experiments, quantum oscillations of the surface electrons have been observed 
in Bi$_2$Te$_3$~\cite{Qu}, and in (Bi,Sb)Se$_3$~\cite{Analytis}.  
The Quantum Hall Effect was also observed in a thick film of strained HgTe~\cite{Molenkamp}.
However, in the Bi-based materials, progress has been slowed by the small surface conductance 
$G^s$ relative to the bulk term $G^b$. We report measurements on 
crystals of Bi$_2$Te$_2$Se in which $G^s/G^b\sim$1 and SdH 
oscillations with large amplitudes are observed at high fields.
By tracking the Landau Level (LL) extrema towards the quantum limit, 
we observe directly the $\frac12$-shift that distinguishes the Dirac spectrum from the Schr\"{o}dinger 
case. Our results address the question whether the spin-Zeeman energy affects the LL sequence in the quantum limit.

Landau quantization of the surface Dirac spectrum 
was previously observed in Bi$_2$Se$_3$ by STM~\cite{Hanaguri,Xue10}. 
Nonetheless, high-$B$ transport experiments to 
approach the quantum limit are important to search for novel states.
In addition, accurate determination of the $\frac12$-shift associated with the Berry phase
provides the best test for whether the SdH oscillations arise from surface topological states
or bulk states 
(this requires a large $B$ to reach the $n$ = 1 LL).

In a magnetic field $\bf B$ normal to the surface,
the Dirac states are quantized into Landau Levels (LLs). 
As $B$ is increased, sequential emptying of the LLs 
leads to oscillations in $G^s$. We follow the customary practice of defining
the ``index field'' $B_n$ as the field at which the Fermi energy $E_F$ lies between two LLs,
i.e. at the minima in $G^s$ (see Sec. \ref{maxmin}). A plot of 
the integers $n$ vs. $1/B_n$ gives a nominally straight line with slope equal to
the FS cross-section $S_F$. 

Our interest is in the limit
$1/B_n\to 0$. In the Schr\"{o}dinger case, there are $n$ filled LLs below $E_F$ when the field equals $B_n$
(as defined).
By contrast, in the Dirac case, we have 
$n+\frac12$ filled LLs between $E_F$ and the Dirac point (at $E$ = 0). The important additional
$\frac12$ arises because the conduction band and the valence band
each contributes half of the states that make up the $n$ = 0 LL. Hence, as $1/B\to 0$, 
the plot of $1/B_n$ vs. $n$ intercepts  
the $n$-axis at the value $\gamma = -\frac12$ for the Dirac case, whereas the intercept $\gamma$ = 0
(mod 1) in the Schr\"{o}dinger case. The $\frac12$-shift was 
experimentally verified for the Dirac spectrum in graphene, and expressed 
equivalently as a Berry-phase $\pi$-shift~\cite{Kim}.

\section{Resistivity maxima or minima?}\label{maxmin}
The index field $B_n$ clearly plays the key role in pinning down the -$\frac12$ shift in the index plot. 
Here we wish to discuss the question of determining $B_n$ when surface and bulk carriers co-exist~\cite{Fu}. 
In the bismuth-based systems (and other 3D topological insulators), 
the two-dimensional electron gas (2DEG) on the surface is in intimate contact 
with bulk electrons which conduct a significant fraction of the 
applied current. By contrast, the entire current is carried by the 2DEG in graphene 
and GaAs heterostructures. When $E_F$ falls between adjacent LLs in the QHE regime of graphene, 
both the 2D conductance $G_s$ and resistance $R_{xx}$ attain 
a deep minimum (this follows from $R_{yx}\gg R_{xx}$).

However, when a large, parallel 
bulk conduction channel exists (the case here), the observed conductance matrix
is the sum 
\be
G_{ij} = G^s_{ij} + G^b_{ij},
\label{eq:G}
\ee
where $G^b_{ij}$ is the bulk conductance matrix. As the mobility of the 
bulk carriers $\mu_b$ is very low (50 cm$^2$/Vs), bulk SdH oscillations are not observable 
even at 45 T. The additivity of the conductances in Eq. \ref{eq:G} implies that the index fields still
correspond to minima in $G_{xx}$. However, because the bulk $G^b_{xx}$ is dominant,
the observed resistance now attains maxima at $B_n$ (i.e. $R_{xx} = G_{xx}/[G_{xx}^2+G_{xy}^2]\sim 1/G_{xx}$).
We find that it is least confusing to work with $G_{ij}$ because its components are additive.
The results reported here provide an experimental
verification of this point.

In many experiments, however, the Hall response is not available. 
One may still use the SdH oscillations in the resistance $R_{xx}$, 
provided $B_n$ is identified with its \emph{maxima}. If the wrong choice is made
(identifying $B_n$ with minima in $R_{xx}$), a spurious -$\frac12$ intercept will
appear for carriers with a Schr\"{o}dinger dispersion.

\begin{figure}[t]
\includegraphics[width=9 cm]{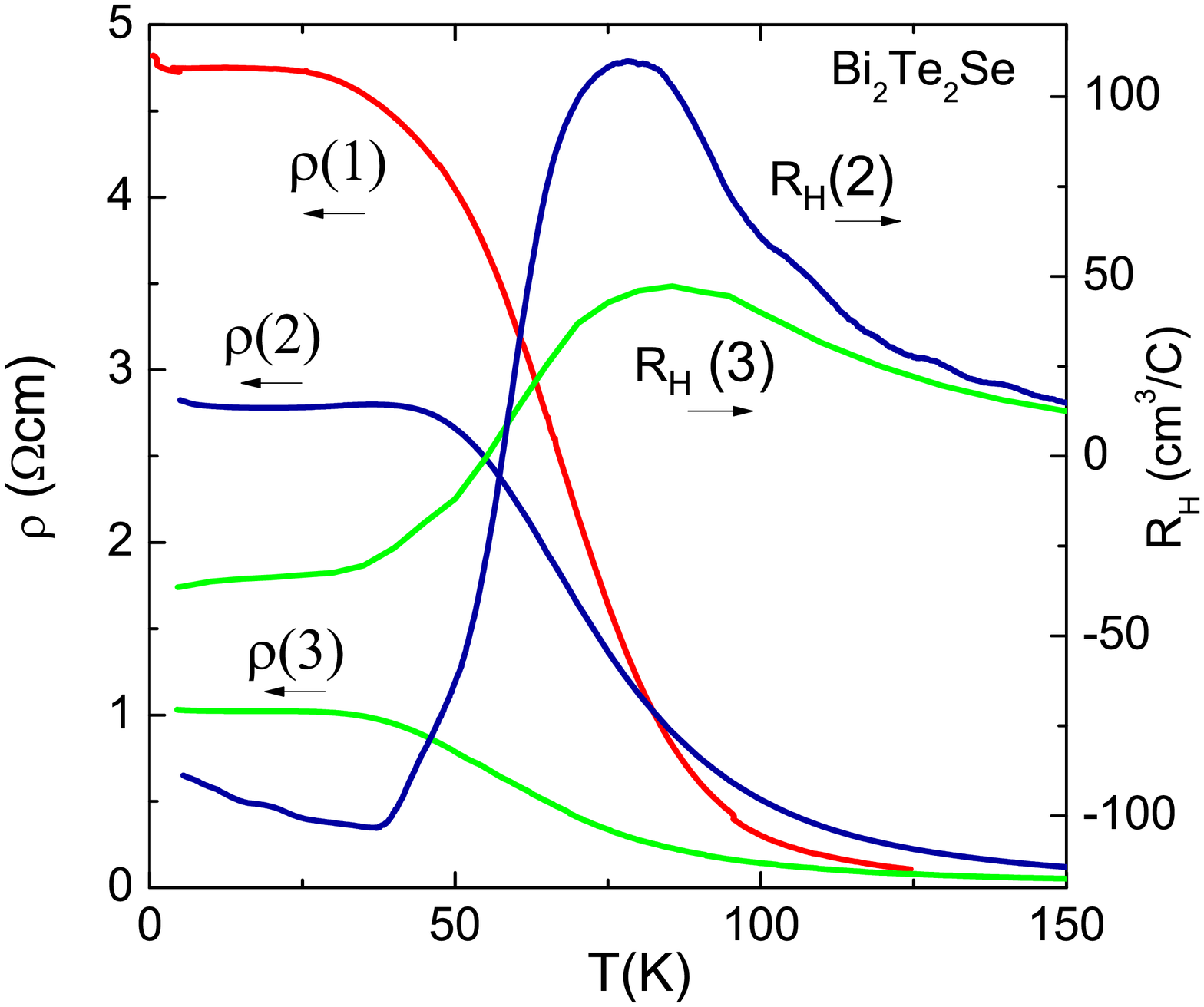}
\caption{\label{figRvsT} (color online)
A representative set of the observed resistivity $\rho$ and Hall coefficient $R_H$ vs. $T$ in Bi$_2$Te$_2$Se (samples identified
by the numbers). The magnitudes of
$\rho$ and $R_H$ at 4 K vary considerably between annealed samples.
In all samples, the carriers are predominantly $n$-type at 4 K. The change in sign of $R_H$ 
near 56 K reflects the thermal activation of bulk hole carriers across a gap of 50 mV.
The largest SdH amplitudes are observed in samples with $\rho >$ 4 $\Omega$cm at 4 K.
}
\efig

A second issue we address is the strength of the Zeeman energy.
Strict particle-hole symmetry implies that it is unshifted in energy. 
On the other hand, a large Zeeman energy $g\mu_BB$ may lead to high-field distortion of the
SdH period ($g$ is the surface Lande g-factor
and $\mu_B$ the Bohr magneton).
The in-field STM experiments ~\cite{Hanaguri,Xue10} have shown that the $n$ = 0 LL is unshifted up to 11 Tesla. 
This test can be extended to much larger $B$ in transport experiments, but early SdH experiments had limited
resolution~\cite{Qu,Analytis}. Values of $g$ as large as 76 have been inferred from low-field
SdH oscillations in Bi$_2$Te$_2$Se~\cite{Taskin}.

\begin{figure}[t]
\includegraphics[width=8 cm]{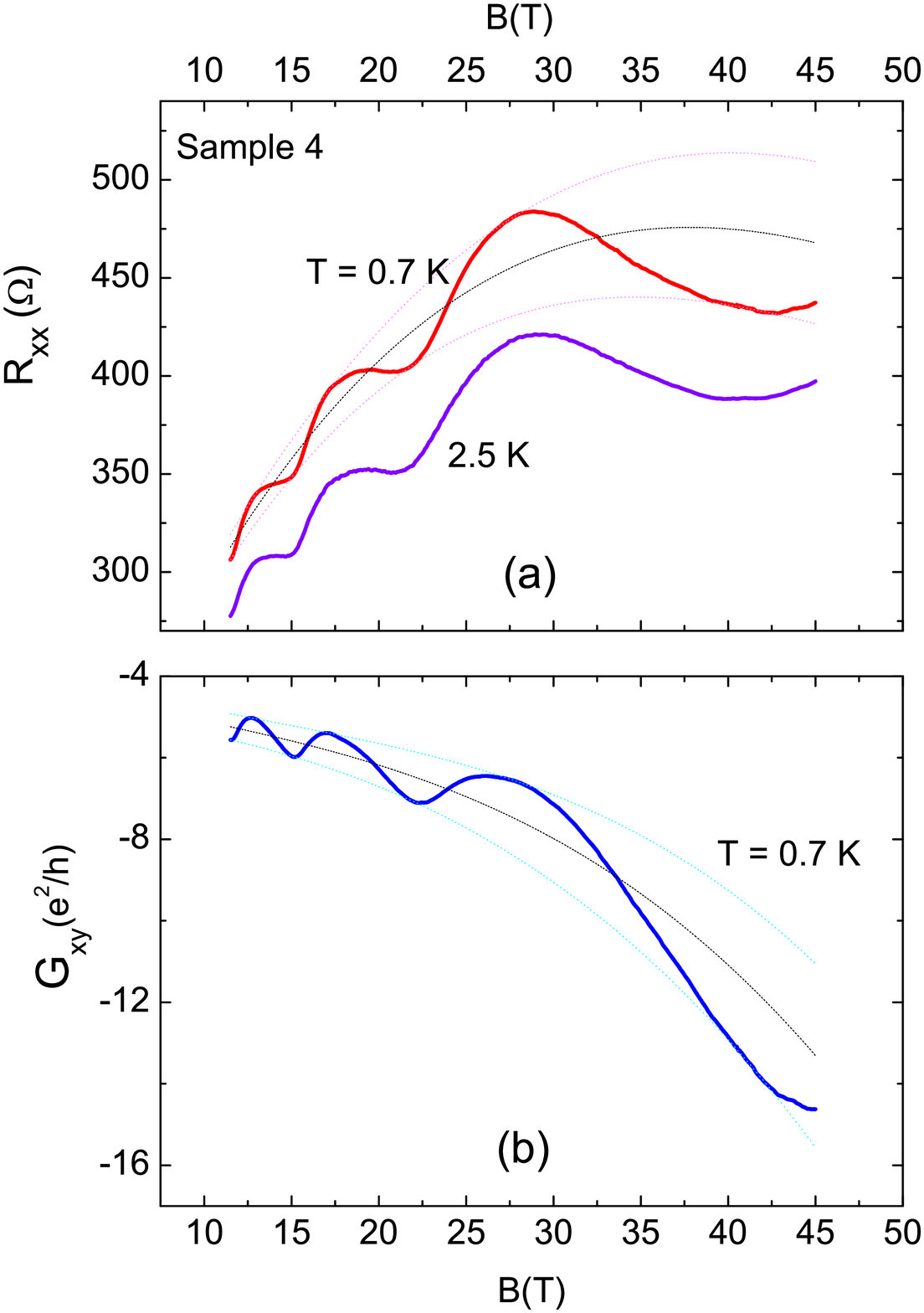}
\caption{\label{figRvsB} (color online)
The resistance (per square) $R_{xx}$ and Hall conductance $G_{xy}$ in Bi$_2$Te$_2$Se (Sample 4).
Panel (a) shows the SdH oscillations in $R_{xx}$ vs. $B$ at fields above 11 T at $T$ = 0.7 and 2.5 K.
At 40 T, the peak-to-peak amplitude is 17 $\%$ of the observed resistance.  The Hall conductance 
$G_{xy}$ at 0.7 K is plotted in Panel (b). In both panels, the envelope is the smooth curve passing through
the extrema points. The background curve (dashed curve)
is determined as the average of the envelope curves.  
}
\end{figure}

\section{Experimental details}
The large density of Se vacancies (electron donors) in Bi$_2$Se$_3$ leads to an $n$-type semi-metal with
a sizeable carrier 
density ($n_b\sim10^{18}$ cm$^{-3}$). By contrast, as-grown crystals of Bi$_2$Te$_3$ are $p$-type because of
Te-Bi exchange defects. In the hybrid material Bi$_2$Te$_2$Se, the Se ions occupy the innermost
layer in each quintuplet layer. This appears to suppress both vacancy formation
and Te-Bi exchange defects. Two groups have found that surface
SdH oscillations are observed in
$n$-type crystals with greatly reduced $n_b$~\cite{Ando10,Xiong}.
Details of the crystal growth for our samples appear in Ref.~\cite{Jia}.

Even in carefully annealed crystals, large variations in the values of $n_b$ and the observed
resistivity $\rho$ are found~\cite{Jia}. Figure \ref{figRvsT} shows traces of $\rho$ vs. $T$
for a representative set (Samples 1, 2 and 3). At 4 K, $\rho$ varies from 1 to 6 $\Omega$cm.
Although all these samples exhibit SdH oscillations, the amplitudes are largest when
$\rho > 4\;\Omega$cm at 4 K.

As shown, the Hall coefficient $R_H$ changes from
$p$ to $n$-type as $T$ decreases near 56 K. 
We have found~\cite{Luo} that the Hall behavior results from the thermal activation 
of holes into the bulk valence band across a ``transport'' gap $\Delta_T\sim$ 50 mV. 
Previously, we showed~\cite{Xiong} that the surface 
conductance $G^s$ in Bi$_2$Te$_2$Se involves carriers with a high mobility $\mu_s$ of 2,800 cm$^2$/Vs,
whereas the residual bulk conductance $G^b$ (from an impurity band) involves $n$-type carriers with much smaller
mobility ($\mu_b\sim$ 50 cm$^2$/Vs). The magnitudes of $G^s$ inferred from $k_F$ and $\mu_s$
confirm that the SdH oscillations are from surface states. 
Ando's group has shown in field-tilt experiments that the SdH period is consistent
with surface states~\cite{Ando10}.
Helical surface states in an isolated Dirac band have been observed by spin-resolved ARPES~\cite{Suyang}.

The large variation in $\rho$ may be understood by estimating the 
number defects. If we assume that each defect (either Se vacancies or Te-Bi exchanges) contributes a carrier, 
the observed $n_b$ (3$\times 10^{16}$ cm$^{-3}$ in Samples 1 and 2) corresponds to a defect density
of a few parts in $10^5$~\cite{Jia}. This stringent constraint implies that fluctuations at this level
lead to pronounced variations in $n_b$ and $\rho$.
Even in optimally annealed
crystals, separate portions of an exposed 
surface can display different $\rho$-$T$ profiles.
In addition, aging of the surface results in 
a gradual decrease in the amplitude of the surface quantum oscillations
with time (roughly by a factor of 2 over a few weeks for crystals 
sealed in Ar atmostphere and stored in dry ice). These factors are problematical 
for high-field transport experiments.

To improve the odds, we cleaved crystals $\sim$30 minutes before loading 
the high-field cryostat. Each crystal was contacted by 3 pairs of leads so that both the
resistance tensor $R_{ij}$ can be measured over distinct segments. 
Because the 45-Tesla field cannot be reversed, we employed the 
reciprocity technique of Ref.~\cite{Sample} to extract both $R_{xx}$ and $R_{yx}$.

\begin{figure}[t]
\includegraphics[width=8 cm]{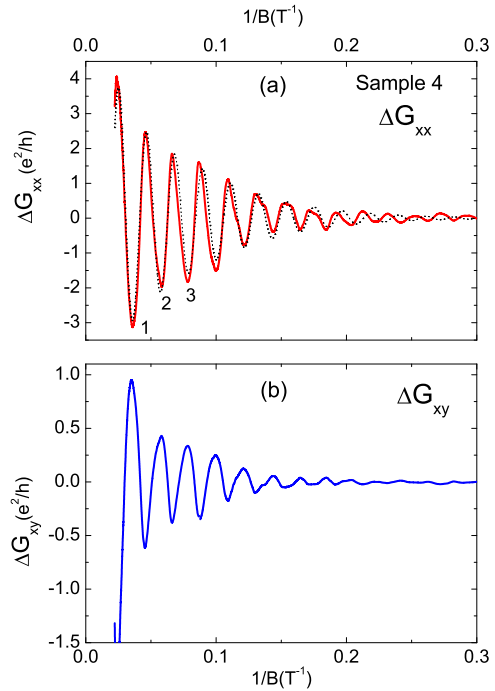}
\caption{\label{figG} (color online)
The oscillatory component of the conductance $\Delta G_{xx}$ (Panel a) and the 
Hall conductance $\Delta G_{xy}$ (Panel b) in Sample 4 plotted against $1/B$ ($T$ = 0.7 K). The two quantities
are normalized to $e^2/h$. 
The fit of the oscillations (see Fig. \ref{figfit}) yields a 
surface mobility of 3,200$\pm$300 cm$^2$/Vs, with $k_F\ell$ = 30. In Sample 4, 
$G^s$ accounts for $\sim 19\%$ of the total conductance at 4 K. Note the phase shift at low $B$.
The LL indices $n$ = 1,2,3 are indicated for the 
minima of $\Delta G_{xx}$. 
}
\end{figure}


\section{Quantum oscillations}
We report measurements to fields of 45 T in Samples 1 and 4
(in which $R_H$ = -137 and -52 cm$^3$/C, respectively, at 4 K). 
The large, well-resolved SdH oscillations in these samples provide an opportunity
to investigate the specific issues in the quantum limit. As shown in Fig. \ref{figRvsB}a, 
the peak-to-peak SdH amplitude in the resistance $R_{xx}$ in Sample 4 grows with $B$
until it accounts for $\sim$17$\%$ of the total resistance. Because conductances
are additive, it is expedient to convert $R_{ij}$ to the conductance
$G_{xx} = R_{xx}/[R_{xx}^2+R_{yx}^2]$ and the Hall conductance $G_{xy}= R_{yx}/[R_{xx}^2+R_{yx}^2]$.
$G_{xy}$ is plotted in Fig. \ref{figRvsB}b.
Using the envelope of the oscillations (faint curves), 
we locate the midpoint between adjacent extrema to define the background.

After removing the background, we isolate the oscillatory components $\Delta G_{xx}$ and $\Delta G_{xy}$
which we plot versus $1/B$ in Fig. \ref{figG}. The conductance $\Delta G_{xx}$ 
and Hall conductance $\Delta G_{xy}$ are plotted in Panels (a) and (b), respectively
(both normalized to the quantum of conductance $e^2/h$).
The fit of the oscillations (see Sec. \ref{fit}) yields a surface mobility of 3,200 cm$^2$/Vs
and a metallicity parameter $k_F\ell$ = 30. The interesting phase shift
apparent at low $B$ is discussed later.

\begin{figure}[t]
\includegraphics[width=9.5cm]{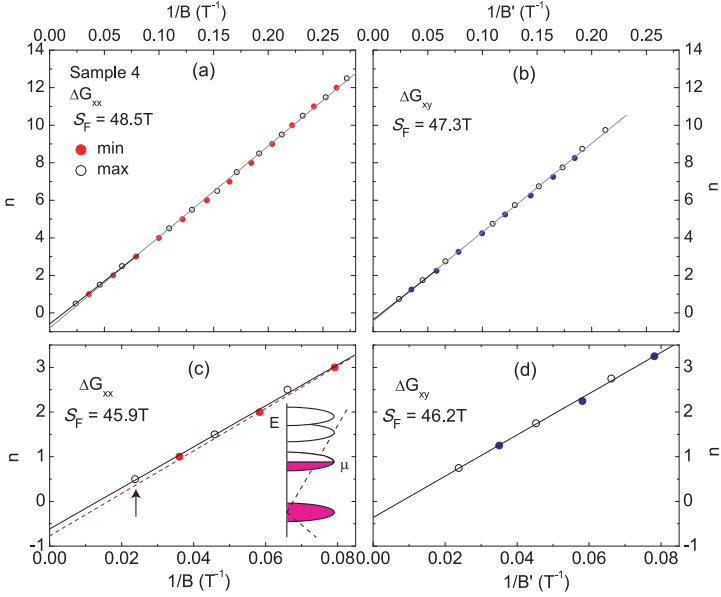}
\caption{\label{figindex} (color online)
The index plots of $1/B_n$ vs. the integers $n$ in Sample 4. In Panel (a), $B_n$ is obtained from
the minima of $\Delta G_{xx}$. In Panel (b), the index field $B'_n$ is inferred from the minima of
$-\Delta G_{xy}$. $B'_n$ is plotted against $n+\frac14$, where the $\frac14$ shift arises because 
the minima in $d\Delta G_{xy}/dB$
align with the minima in $\Delta G_{xx}$. We expand the scale in Panels (c) and (d) to
show the intercepts more clearly. 
In Panel (c), the solid straight line is the best fit to the extrema fields for $n\le$3. The
dashed line is the best fit to all the extrema field shown in Panel (a).
The sketch shows $E_F$ in relation to the filled LLs (solid color) in the Dirac spectrum when
$B$ = 42.0 T (arrow). 
}
\end{figure}


Figure \ref{figindex}a plots the minima 
of $\Delta G_{xx}$ versus $n$ (solid circles). In addition, 
the maxima of $\Delta G_{xx}$ have been plotted as open circles (shifted by $\frac12$).
The best-fit straight line gives a Fermi cross-section area $S_F$ of 48.5 T.
A similar plot based on the extrema of the Hall conductance $\Delta G_{xy}$ 
is shown in Fig. \ref{figindex}b. The minima in $-\Delta G_{xy}$ correspond
to $n+\frac14$, since the derivative $-d\Delta G_{xy}/dB$ has minima at $n$.
The value of $S_F$ found from $\Delta G_{xy}$ (47.3 T) is consistent with 
the previous value within our resolution. 
The values of $n$ = 1,2,3 at the minima of $\Delta G_{xx}$ are noted in Fig. \ref{figG}a.

In order to fix the intercept $\gamma$, we expand the scale in Fig. \ref{figindex}c.
The best-fit straight line passing through the six extrema of $\Delta G_{xx}$ 
intercepts the $n$-axis at the value $\gamma$ = -0.61$\pm$0.03. Similarly, the high-field
extrema of $\Delta G_{xy}$ are plotted in Fig. \ref{figindex}d.  
The intercept for the best-fit line occurs at $\gamma$ = -0.37$\pm$0.03. 
Within our uncertainties, these intercepts are significantly closer to the ideal value $\gamma=-\frac12$ than
0 or 1. Hence, the high-field results provide transport evidence
for a Dirac spectrum for the surface states.

Although we do not observe quantized Hall steps in Fig. \ref{figG}b (the oscillatory component
rides on a large tilted background contribution from the bulk Hall current), it is interesting that the peak-to-peak
amplitude swing of $\Delta G_{xy}$ is $\sim$0.8 $e^2/h$ per surface for $n=1$, which is of the order
of the quantized Hall conductance value.

In Sample 1, the amplitudes of the observed SdH oscillations are considerably
weaker (Fig. \ref{figG1}a). The index plot of $1/B_n$ vs. $n$ fits a straight line
that intercepts the $n$-axis at $\gamma$ = -0.45$\pm$0.02, again consistent with
a Dirac spectrum.

The expanded plot shows why intense fields are needed to fix $\gamma$ reliably. 
By accessing the $n=\frac12$ index at 45 T (Figs. \ref{figindex}c,d), 
we have reduced considerably the ``spread'' of intercepts caused by 
the measurement uncertainties: an intercept $\gamma = 0$ may be safely excluded.
A more subtle point is the slight curvature of the index plot.
In Fig. \ref{figindex}c, if we extrapolate the best-fit line (dashed) using the total 
data set from 3 to 45 T, its intercept yields -0.78, nearly exactly between -1 and
-$\frac12$. By contrast, the best-fit line (bold) to the high-field extrema
for $n\le$3 yields an intercept (-0.61) closer to -$\frac12$. This implies that the index curve
$1/B_n$ vs. $n$ develops a slight curvature in intense fields. (The curvature accounts
for the low-$B$ phase shift apparent in the single-frequency fit in Fig. \ref{figG}a.)

\begin{figure}[t]
\includegraphics[width=8 cm]{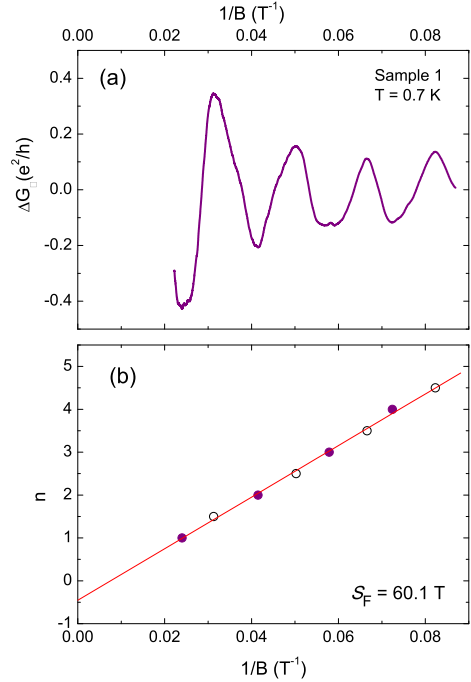}
\caption{\label{figG1} (color online)
The oscillatory component $\Delta G_{xx}$ vs. $1/B$ (Panel a) and the 
index plot of $1/B_n$ vs. $n$ (Panel b) in Sample 1. The intercept $\gamma$ of the best-fit line 
is -0.45$\pm$0.02.}
\end{figure}


A possible cause of curvature is the spin-Zeeman energy. When that is included, the Hamiltonian is
\be
H = v_F {\bf \hat{n}}\cdot \bm{\sigma\times\pi}- \frac{g\mu_B}{2} {\bf B}\cdot\bm{\sigma}
\label{eq:H}
\ee
where $\bf \hat{n}$ the unit vector
normal to the surface. $\bm\sigma$ are the spin Pauli matrices, and $\bm{\pi} = {\bf p}-e{\bf A}$ is the momentum $\bf p$ of the
electron in a vector potential ${\bf A}$. The LL energy is given by 
\be 
E_n = \pm\sqrt{2n\hbar v_F^2 eB + (g\mu_B B/2)^2}.
\label{eq:En}
\ee
The energy of the $n=0$ LL increases linearly with $B$ instead of being unshifted.
For a large $g$, the plot of $1/B_n$ vs. $n$ will deviate from a straight line as $1/B\to$ 0. In our experiment, we have
tracked the LLs to $n$ = 1. The weak deviation from a straight line in Fig. \ref{figindex}c) 
is inconsistent with values of $g$ substantially larger than 2. More importantly, however, the
observed deviation is opposite in sign to that predicted by Eq. \ref{eq:En}. 
As we do not see evidence for a deviation caused by a large $g$-factor, we conclude that the 
the $g$ factor of the surface states in Bi$_2$Te$_2$Se are not significantly greater than 2 in the quantum limit. 

\section{Surface carrier mobility}\label{fit}
In general, it is very difficult to separate $G^s$ from $G^b$ reliably even at $B$ = 0.  
Shubnikov de-Haas (SdH) oscillations -- when measured with sufficient resolution -- 
provide a powerful way to tease out the surface conductance.
Analysis of the SdH amplitude vs. $B$ yields the scattering rate and
the surface mobility $\mu_s$ (equivalently the mean-free-path $\ell$). Also,
the period of the oscillations yields $k_F$. 
With $\mu_s$ and $k_F$ known, we then obtain the zero-$B$ value of $G^s_{xx}\equiv G^s$ using
\be
G^s = (e^2/h)k_F\ell.
\label{eq:Gs}
\ee

To focus on the SdH oscillations, we first determine the envelope curves
passing through the extrema of the oscillations 
as explained in Fig. 2 of the main text. The oscillatory component $\Delta G_{xx}$ is
obtained by subtracting from $G_{xx}$ the background, defined as the 
curve lying between the envelope curves.
(We remark that $\Delta G_{xx}$ does not account for all of the
surface conductance. By construction, its field-averaged value $\langle\Delta G_{xx}\rangle_B$
vanishes. Hence we must have $\Delta G_{xx}<G^s_{xx}$.)


\bfig[t]            
\incl[width=9cm]{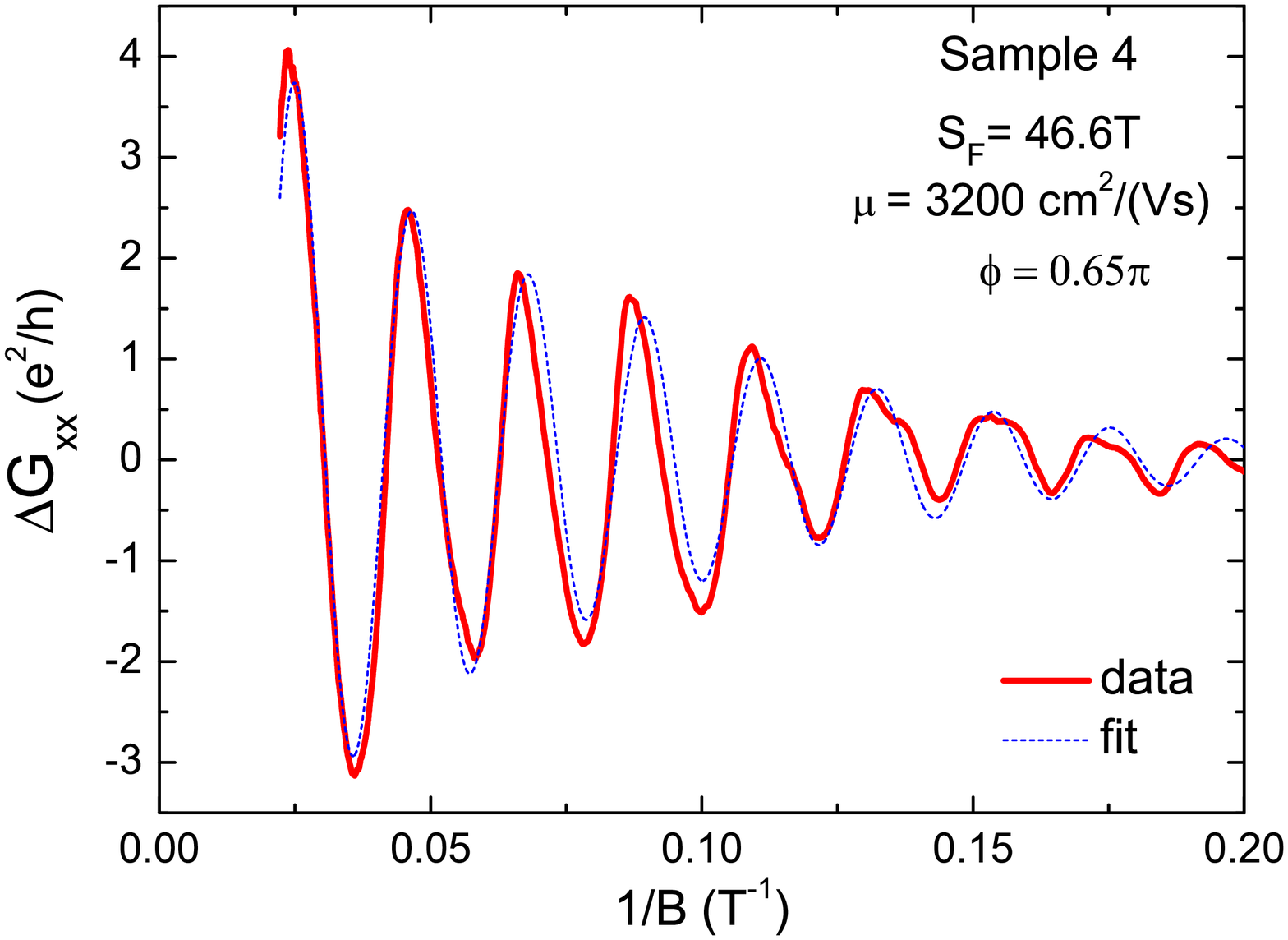} 
\caption{\label{figfit} (Color online) 
The oscillatory component of the conductance $\Delta G_{xx}$ in Sample 4 at 0.7 K
(solid curve) and the fit to Eq. \ref{eq:sdh} using only one frequency (dashed curve). 
}
\efig

To fit the oscillatory component $\Delta G_{xx}$, we employed
the standard Lifshitz-Kosevich expression~\cite{Roth} 
\be
\frac{\Delta G_{xx}}{G_{xx}} = \left(\frac{\hbar\omega_c}{2E_F}\right)^{\frac12}
\frac{\lambda}{\sinh\lambda} e^{-\lambda_D}\cos
\left[\frac{2\pi E_F}{\hbar\omega_c}+\varphi\right],
\label{eq:sdh}
\ee
with $\lambda = 2\pi^2k_BT/\hbar\omega_c$ and $\lambda_D = 2\pi^2k_BT_D/\hbar\omega_c$,
where $\omega_c$ is the cyclotron frequency and 
the Dingle temperature is given by $T_D = \hbar/(2\pi k_B\tau)$,
with $\tau$ the lifetime. For 2D systems, we may write the SdH frequency as
$2\pi E_FB/(\hbar\omega_c)$, which simplifies to $4\pi^2\hbar n_s/e$, with the 2D
carrier density $n_s = k_F^2/4\pi$ (per spin).  
Equation \ref{eq:sdh} may be employed in a Dirac system
if we write the cyclotron mass as $m_c = E/v_F^2$.

As shown in Fig. \ref{figfit}, we obtain a reasonably close fit to the observed oscillations (bold curve)
using just one frequency. The optimal fit yields for the 3 adjustable parameters
the values $k_F$ = 0.038 \AA$^{-1}$, $\varphi$ = 0.65$\pi$ and
$T_D$ = 8.5$\pm$1.5 K, which implies a surface 
mean-free-path $\ell$ = 79$\pm$8 nm and mobility $\mu_s = e\ell/\hbar k_F$ = 3,200$\pm$300 cm$^2$/Vs.
The metallicity parameter $k_F\ell$ equals 30. 
We estimate that, in Sample 4 at $B$=0, $G^s$ accounts for $\sim 19\%$ of the
total observed conductance.
These values are similar to those obtained in an earlier
sample, which had a slightly larger $k_F$ (0.047 \AA$^{-1}$)~\cite{Xiong}.

The mobility provides a strong, quantitative argument that the SdH oscillations
arise from surface states.  Suppose for the sake of argument that the oscillations arise 
from bulk states. The SdH period is then to be identified with a 3D Fermi sphere 
of radius $k_F$ = 0.038 \AA$^{-1}$, or a 3D carrier density of 1.86$\times 10^{18}$ cm$^{-3}$.  With
this density, the inferred mobility gives a 3D resistivity $\rho_b \sim$ 1.1 m$\Omega$cm at 4 K.
Instead we measure $\rho$ to be 5 $\Omega$cm. 
The large discrepancy (factor of 4,500) firmly precludes a 
bulk origin for the SdH oscillations.

\section{Conclusions}
The Dirac-like topological surface states detected in ARPES and STM experiments
present a host of new opportunities for transport experiments especially in high magnetic fields.
In bulk crystals, the presence of bulk carriers complicate transport studies. 
As shown here, quantum oscillations provide a powerful way to isolate the
surface carriers and to determine their mobility and $k_F\ell$. The index plot 
of the integers $n$ versus $1/B_n$ can be used to confirm the 
$\pi$-shift associated with the Berry phase of the surface electrons, which leads to
an intercept -$\frac12$ in the limit $1/B\to 0$. To access LLs at $n$ = 1 (or lower), 
we have employed fields up to 45 T. The results in Figs. \ref{figindex} and \ref{figG1} 
provide direct confirmation of the existence of the -$\frac12$ intercept expected from
a Dirac disperion.

The resolution attained here provides experimental verification 
of the point that the -$\frac12$ intercept is observed only when $B_n$ is 
identified with minima in $G_{xx}$ or maxima in $R_{xx}$. (For contrast, we note a recent 
report~\cite{Veldhorst} in which a -$\frac12$ intercept was obtained in high-$B$ measurements on 
exfoliated crystals of Bi$_2$Te$_3$.
However, because $B_n$ was inferred from minima in the resistivity, it seems 
that the -$\frac12$ intercept actually implies a Berry phase that is zero, consistent with SdH oscillations from conventional bulk carriers. 
) 

The linearity of the index plot in Figs. \ref{figindex} and \ref{figG1} show that the Lande
$g$-factor is small ($g\sim$2). The $n$ = 0 LL is unshifted even at 45 T, consistent with 
STM experiments taken at 11 T~\cite{Hanaguri,Xue10}.

Finally, we comment on the results in the large-$B$ limit.
In Fig. \ref{figG}a, the last maximum in $\Delta G_{xx}$ (at $B\simeq$ 40 T) corresponds to
$n$ = $\frac12$ (see arrow in the index plot in Fig. \ref{figindex}c). At this field, the
Fermi energy $E_F$ is aligned with the center of the $n$ = 1 LL, as sketched in the
inset in Fig. \ref{figindex}c. In our indexing scheme, there is 1 filled
LL between $E_F$ and the Dirac Point, with $\frac12$ of the filled states 
from the unshifted LL at the Dirac Point). Hence, these results provide 
rather firm evidence 
for this $\frac12$-shift in the limit $1/B\to 0$. 
As the inset in Fig. \ref{figindex}c implies, the interesting
states in the $n$ = 0 LL in Sample 4 become experimentally accessible in fields higher than 45 T.

\vspace{3mm}
We acknowledge valuable discussions with Liang Fu, Xiao Liang Qi and Joel Moore. 
The research is supported by the US National Science Foundation (grant DMR 0819860)
and the Army Research Office (ARO W911NF-11-1-0379).
Sample growth and characterization were supported by an award from the 
Defense Advanced Research Projects Agency under SPAWAR Grant No.: N66001-11-1-4110. 
The experiments were performed at the 
National High Magnetic Field Laboratory, which is supported by NSF 
Cooperative Agreement No. DMR-084173, by the State of Florida, and by the Department of Energy. 
YKL acknowledges support from the China Scholarship Council (CSC).\\

$^*$\emph{Current address of YKL}: Department of Physics, Zhejiang University, Hangzhou, China.


%

%

\begin{thebibliography}{99}
\bibitem{Fu} We are indebted to Liang Fu for clarifying this point.
\end{thebibliography}
\end{document}